\documentclass[preprint,12]{aastex}
\begin{document}

\title{The Top Ten {\em Spitzer\/} YSOs in 30 Doradus} 

\author{Nolan R.\ Walborn}
\affil{Space Telescope Science Institute,\altaffilmark{1} 
3700 San Martin Drive, Baltimore, MD 21218}
\email{walborn@stsci.edu}

\author{Rodolfo H.\ Barb\'a\altaffilmark{2}}
\affil{Departamento de F\'{\i}sica, Universidad de La Serena, 
Cisternas 1200 Norte, La Serena, Chile}
\email{rbarba@dfuls.cl}

\and

\author{Marta M.\ Sewi{\l}o\altaffilmark{3}}
\affil{Department of Physics \& Astronomy, The Johns Hopkins University,\\ 
3400 N.~Charles Street, Baltimore, MD 21218}
\email{mmsewilo@pha.jhu.edu}

\altaffiltext{1}{Operated by AURA, Inc., under NASA contract NAS5-26555.}

\altaffiltext{2}{Also at Instituto de Ciencias Astron\'omicas de la Tierra 
y del Espacio (ICATE--CONICET), Avenida Espa\~na 1512 Sur, J5402DSP, San Juan, 
Argentina.}

\altaffiltext{3}{Also at Space Science Institute, 4750 Walnut St., Suite 205, 
Boulder, CO 80301.} 

\begin{abstract}
The most luminous {\it Spitzer\/} point sources in the 30~Doradus triggered 
second generation are investigated coherently in the 3--8~$\micron$ region.  
Remarkable diversity and complexity in their natures are revealed.  Some are also among the brightest \textit{JHK} sources, while others are not.  Several of them are multiple when examined at higher angular resolutions  with {\it HST\/} NICMOS and WFPC2/WFC3 as available, or with VISTA/VMC
otherwise.  One is a dusty compact H~II region near the far northwestern edge of the complex, containing a half dozen bright $I$-band sources.  Three others appear closely associated with luminous WN stars and causal connections are suggested.  Some are in the heads of dust pillars oriented toward R136, as previously discussed from the NICMOS data.  One resides in a compact cluster of much fainter sources, while another appears monolithic at the highest resolutions.  Surprisingly, one is the brighter of the two extended ``mystery spots'' associated with Knot~2 of  Walborn et~al.  Masses are derived from YSO models for unresolved sources and lie in the 10--30~$M_{\sun}$ range.  Further analysis of the IR sources in this unique region will advance understanding of triggered massive star formation, perhaps in some unexpected and unprecedented ways.
\end{abstract}


\section{Introduction}

The first evidence of discrete, current star formation in 30~Doradus (the largest H~II region in the Large Magellanic Cloud and indeed in the Local Group) was found by Hyland et~al.\ (1992), who reported four luminous 
protostars, and by Rubio et~al.\ (1992), who discovered 17 near-IR sources 
arrayed in arcs to the west and north of the central cluster R136.  Prior to that, 30~Dor had been considered an evolved H~II region without major ongoing star formation, although ``extended'' (unresolved) far IR emission had been reported west and northeast of R136 already by Werner et~al.\ (1978).  More extensive subsequent investigations  with increasingly powerful instrumentation have established that in fact, 30~Doradus is a ``two-stage starburst'' (Walborn \& Parker 1992), with substantial triggered star formation primarily at the interfaces between the expanding central cavity and remnant giant molecular clouds to the west and northeast (Walborn \& Blades 1997; Rubio et~al.\ 1998; Johansson et~al.\ 1998; Scowen et~al.\ 1998; Walborn et~al.\ 1999, 2002; Brandner et~al.\ 2001).

The {\it Spitzer Space Telescope\/} has provided the best, unimpeded imaging capability so far at longer IR wavelengths than those readily observed from 
the ground (or from the {\it Hubble Space Telescope\/}).  Extensive lists of sources in the Large Magellanic Cloud observed by the SAGE program (Meixner et~al.\ 2006, see below), including a number within 30~Doradus, have been provided by Whitney et~al.\ (2008) and Gruendl \& Chu (2009).
Specific analysis of 30~Dor with the {\it Spitzer\/} data has concentrated on the diffuse emission (Indebetouw et~al.\ 2009). Here we undertake the first detailed study of the individual and global characteristics of the brightest {\it Spitzer\/} point sources in 30~Dor, in the context of previous work at shorter wavelengths.   

\section{Data}

Our study is based upon the four {\it Spitzer\/}/IRAC bands covering the 
3.6--8.0~$\micron$ region, in which the angular resolution ranges from   
$1\farcs7$ through $2\farcs0$, respectively.  Because of the endemic 
multiplicity of massive stars and the distance of the LMC (at which 1\arcsec\ 
subtends 50,000~AU), the highest available resolutions are essential.  Thus, we also discuss $\sim0\farcs1$ {\it HST\/} NICMOS/WFPC2 (Walborn et~al.\ 
1999, 2002; Brandner et~al.\ 2001) and WFC3 (E.~Sabbi et~al.\ in prep.) data, as well as $\sim1\arcsec$ groundbased Rubio et~al.\ (1998), 2MASS (Skrutskie et~al.\ 2006), IRSF (Kato et~al.\ 2007), and VISTA/VMC (Cioni et~al.\ 2011) data in comparison with {\it Spitzer\/}.  Some further characteristics of the surveys and data used in our analysis follow.  

\subsection{{\em Spitzer\/}/SAGE}

The {\it Spitzer Space Telescope\/} Legacy Program ``Surveying the Agents
of a Galaxy's Evolution'' (SAGE; Meixner et~al.\ 2006) imaged a $7^{\circ} \times 7^{\circ}$ region in the LMC with all the IRAC (3.6, 4.5, 5.8, and 8.0~$\mu$m; Fazio et~al.\ 2004) and MIPS (24, 70, and 160~$\mu$m; Rieke et~al.\ 2004) bands.  The MIPS data are not used here because of their lower angular resolution.  We use the IRAC ``Single Frame + Mosaic Photometry'' Archive.\footnote{See ``The SAGE Data Products Description'' at http://data.spitzer.caltech.edu/popular/sage/20090922\_enhanced/\break documents/.}

\subsection{VISTA/VMC}

The ``VISTA Magellanic Clouds'' survey (VMC; Cioni et~al.\ 2011) is a 
European Southern Observatory (ESO) public survey, which consists of deep
imaging in the three NIR bands $Y$ (1.02~$\mu$m), $J$ (1.25~$\mu$m), and 
$K_{\rm s}$ (2.15~$\mu$m), covering both Magellanic Clouds and surrounding areas, with the VIRCAM (VISTA IR Camera; Dalton et~al.\ 2006) at the ESO VISTA telescope.  The first release of VMC data includes images of the 30 Doradus Region (a ``tile'' of about 1$\rlap.^{\circ}0$ $\times$ 1$\rlap.^{\circ}3$).

We retrieved the 30~Doradus images through the ESO Data Archive Facility.
For our analysis, we used the ``pawprint'' images (six of which form a ``tile''). The released VMC data have been processed through the dedicated pipeline of the VISTA Data Flow System (Irwin et~al.\ 2004, Emerson et~al.\ 2004).
In addition to the images, the VMC survey contains source catalogues, but the aperture photometry they provide includes only a few of our sources and may be unreliable in regions of bright, variable nebulosity.

Accordingly, we performed point-spread-function (PSF) photometry of our
sources in the VMC images, using DAOPHOT (Stetson 1987) software layered in IRAF.\footnote{IRAF is distributed by the National Optical Astronomy Observatory.}  Stars were detected at a minimum $5 \sigma$ level above the background, with special care to isolate faint companions to the targets.  

The VISTA photometric system is tied to the Two Micron All Sky Survey (2MASS; Skrutskie et~al.\ 2006).  To derive the photometric cross-calibrations, 
we have selected 440 2MASS catalogue stars with the lowest photometric uncertainties ($<$10\% in all three bands) and located in areas of faint nebulosity, matching them to our VISTA photometric catalogue. The transformation equations obtained are 
$$J_{\rm 2MASS} - J_{\rm VISTA} = 0.06 \times (J_{\rm VISTA} - K_{\rm VISTA})$$ and
$$K_{\rm 2MASS} - K_{\rm VISTA} = -0.003 \times (J_{\rm VISTA} - K_{\rm VISTA})~~.$$
The coefficients of these equations are very similar to those found by Cioni et~al.\ (2011).  Our transformations are valid for the color range of the 2MASS sources used to calculate them, i.e., $-0.1 < (J-K_{\rm s}) < 4.2$. Table~1a shows that the 30~Dor sources with both $J$ and $K_{\rm s}$ magnitudes
are in the valid range. For sources with only limits on $J$, the uncertainties
in the $K_{\rm s}$ transformation are also very small, due to the small value of
the color-term coefficient.  (The $Y$~magnitudes in Table~1a remain in the VISTA photometric system and are not used in our analysis below.)

\subsection{IRSF}

The Infrared Survey Facility (IRSF) Magellanic Clouds Point Source Survey
(Kato et~al.\ 2007) covered a 40~deg$^2$ area in the LMC in $J$ (1.25~$\mu$m), $H$ (1.63~$\mu$m), and $K_{\rm s}$ (2.14~$\mu$m) bands, with the SIRIUS camera at the South African Astronomical Observatory (SAAO). The average seeing throughout the survey was $1\farcs2$ (FWHM). 

\subsection{Results}

Figures~1 and 2 display {\it Spitzer\/} and VISTA images of the field
investigated at the same scale, with the ten brightest {\it Spitzer\/} sources and some other relevant objects identified.  It is seen that all of them lie within or near the Tarantula filaments to the west and northeast of R136, except for the very luminous and previously little known source S1 to the northwest.  (The hemispheric structure of two-stage starbursts, as discussed by Walborn 2002, is well seen.)  The photometric data taken from the literature, except for VISTA/VMC as discussed above, for these and other IR sources of interest are listed in Table~1a.  The alternate designations are from Hyland et~al.\ (1992) or Rubio et~al.\ (1998), while the corresponding SAGE and IRSF identification numbers are given in Table~1b.  Figure~3 shows that the 30~Doradus sources are among the most luminous YSOs in the LMC, as listed in the caption references.  The 30~Dor sources are discussed individually in detail in the following section, emphasizing their remarkable structures and diversity. 

\begin{figure}
\epsscale{0.8}
\plotone{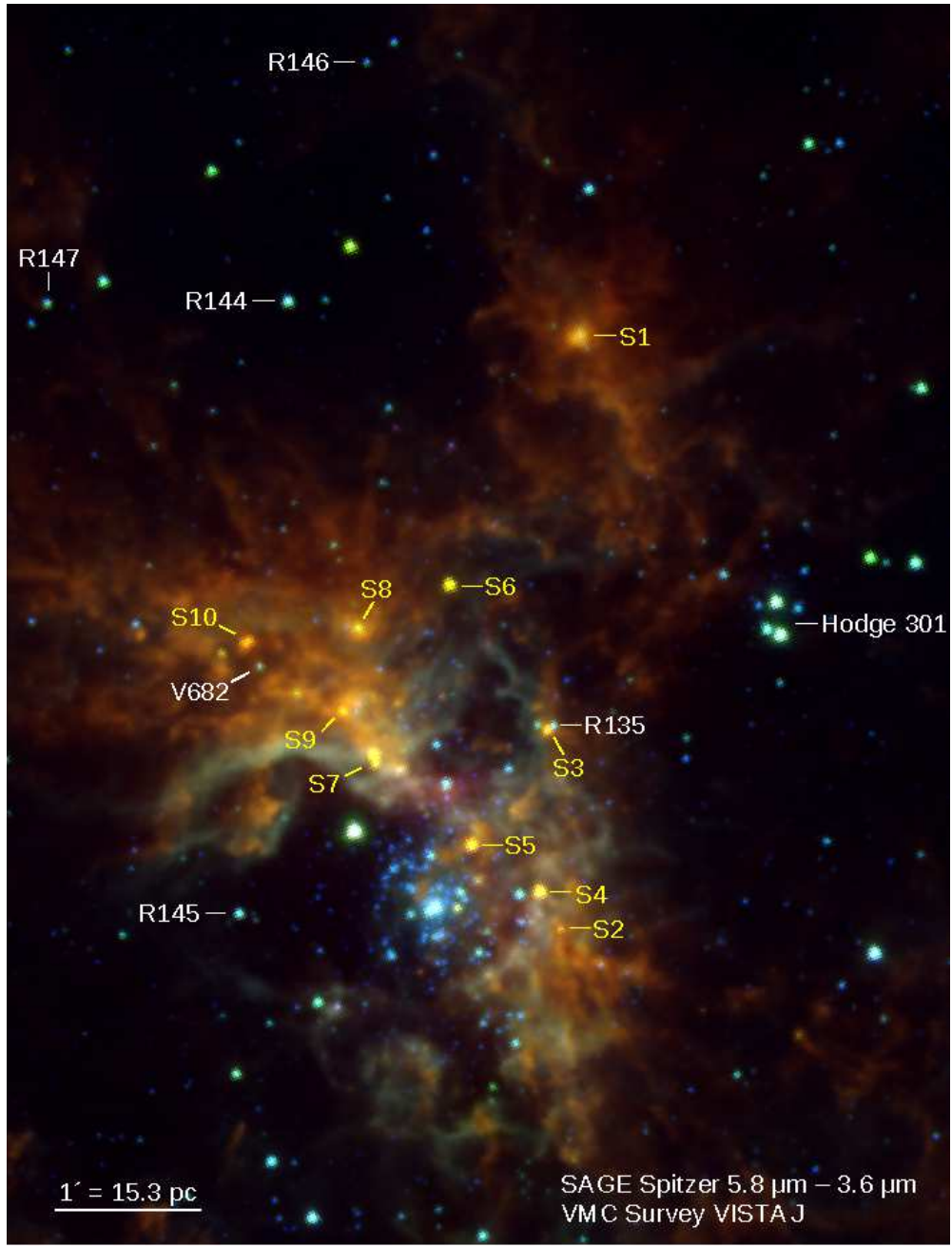}
\caption{\label{fig:fig1} A composite of the full 30~Doradus Nebula field in 
two {\it Spitzer\/} and a VISTA/VMC bands.  North is up and east to the left.  
The top ten IR sources are labeled in yellow~(S), while bright, isolated WN 
stars are in white (R~for Radcliffe, V~for VFTS).  R136 is the bright blue object below center, while the older cluster Hodge 301 with late-type 
supergiants (green) is labeled.  Note the color range of the IR sources with 
this broad baseline.}
\end{figure}

\begin{figure}
\epsscale{0.8}
\plotone{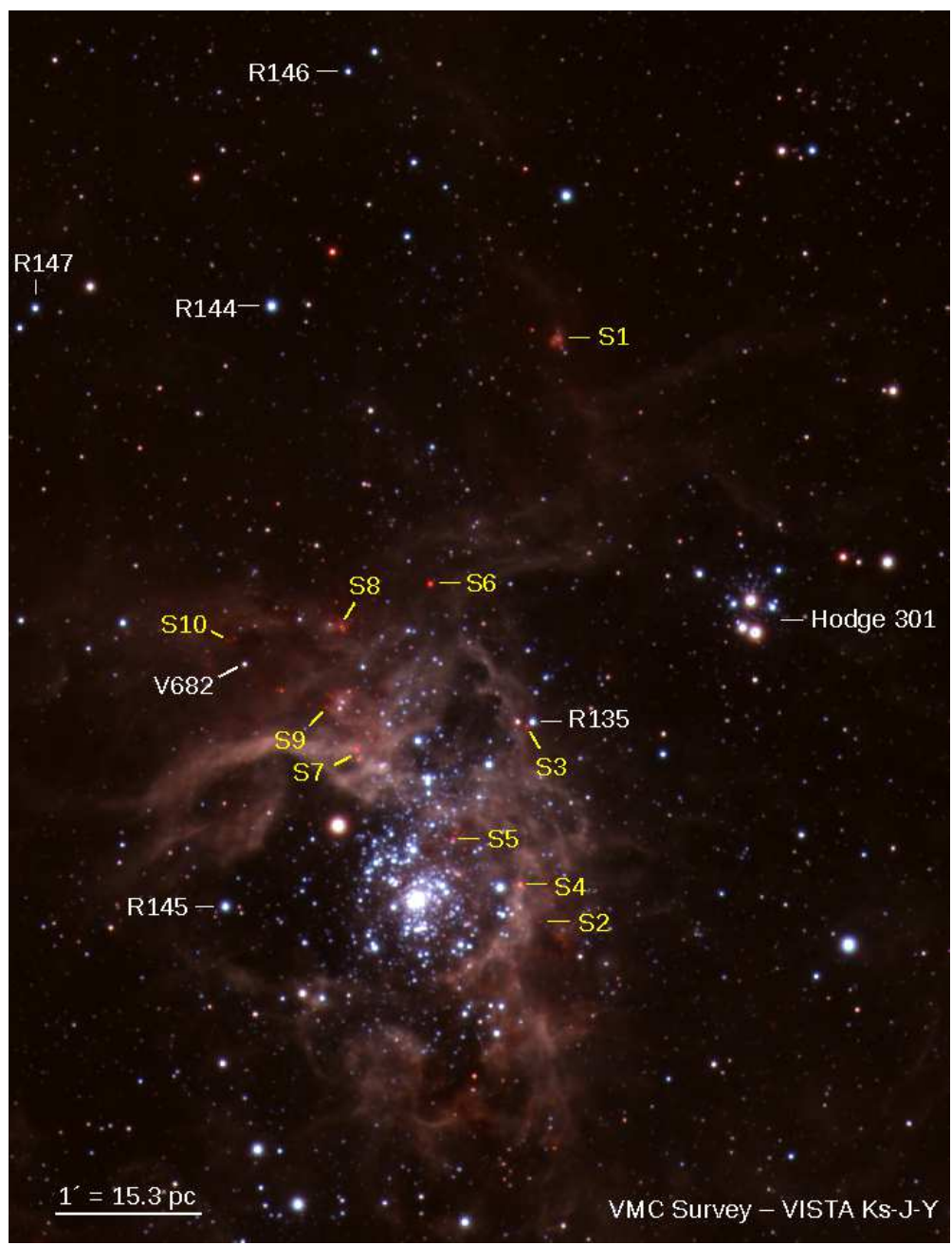}
\caption{\label{fig:fig2} Same as Fig.~1 but entirely in VISTA/VMC bands, 
which have higher resolution.  Note the multiplicity of several IR sources as 
discussed in the text, as well as the reversals of relative magnitudes in the S3 and S10 systems with respect to Fig.~1.}
\end{figure}

\begin{figure}
\epsscale{1.0}
\plotone{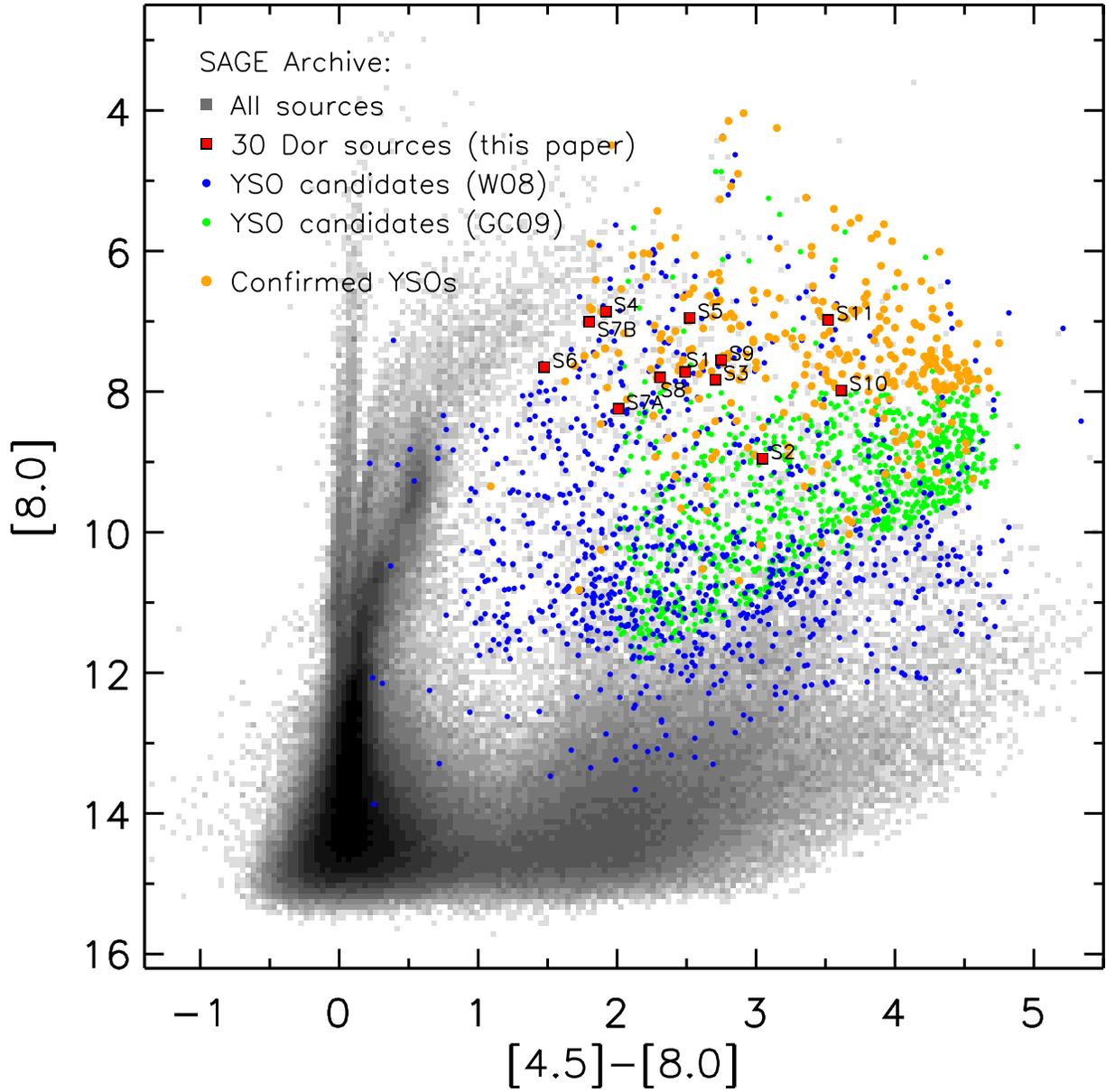}
\caption{\label{fig:fig3} A {\it Spitzer\/} CMD comparing the brightest 30~Dor sources (red squares) with the LMC YSO candidates of Whitney et~al.\ (2008) and Gruendl \& Chu (2009), and the confirmed YSOs of Shimonishi et~al.\ 2008, Oliveira et~al.\ 2009, Seale et~al.\ 2009, and Woods et~al.\ 2011.  The present sources are in the brighter region of the distribution, although several of them are multiple as discussed in the text.}
\end{figure}

\clearpage
\voffset.3in
\hoffset1in
\thispagestyle{empty}
\begin{deluxetable}{@{}llcccccccccccccc@{}}
\tabletypesize{\tiny}
\rotate  
\tablewidth{9.45in}
\tablenum{1a}
\tablecolumns{16}
\tablecaption{Photometry}
\tablehead{
\colhead{{\it Spitzer}} &
\colhead{Alternate} &
\colhead{R.A.(J2000)} &
\colhead{Decl.(J2000)} &
\multicolumn{3}{c}{VISTA} &
\colhead{} &
\multicolumn{3}{c}{IRSF} &
\colhead{} &
\multicolumn{4}{c}{IRAC} \\

\cline{5-7}
\cline{9-11}
\cline{13-16}

\colhead{Source} &
\colhead{IDs} &
\colhead{($^{h~m~s}$)} &
\colhead{($^{\circ~'~''}$)} &
\colhead{$Y$} &
\colhead{$J$} &
\colhead{$K_{s}$} &
\colhead{} &
\colhead{$J$} &
\colhead{$H$} &
\colhead{$K_{s}$} &
\colhead{} &
\colhead{[3.6]} &
\colhead{[4.5]} &
\colhead{[5.8]} &
\colhead{[8.0]} }

\startdata
S1  \dotfill  &               &  05:38:31.62  & --69:02:14.6  &  16.68(0.1)\phn  &    16.10(0.09) &  14.59(0.08)  & &   \nodata     & \nodata       &  \nodata       &&   10.21(0.06)   &   \phn9.37(0.06)   &    \phn7.72(0.06)     &  5.73(0.06) \\
S2  \dotfill  &   IRSW-11     &  05:38:33.09  & --69:06:11.7  &  \nodata     &    18.80(0.4)\phn  &  14.98(0.02)  & &   \nodata     & 16.95(0.09)   &  15.34(0.08)   &&   12.0\phn(0.10)    &   10.56(0.07)  &    \phn8.95(0.07)     &  7.60(0.07) \\
S3  \dotfill  &               &  05:38:34.05  & --69:04:52.2  &  \nodata     & $>$19          &  14.32(0.03)  & &   \nodata     & 16.41(0.05)   &  \nodata       &&   10.5\phn(0.05)    &   \phn9.08(0.04)   &    \phn7.83(0.04)     &  5.88(0.04) \\
S3K \dotfill  &               &  05:38:34.69  & --69:04:50.0  &  14.27(0.01) &    13.37(0.01) &  11.71(0.01)  & &   13.21(0.02) & 12.17(0.01)   &  11.69(0.01)   &&   11.27(0.05)   &   11.29(0.07)  &    10.02(0.21)    &  \nodata    \\
S4  \dotfill  &   IRSW-30;P3  &  05:38:34.60  & --69:05:56.8  &  16.01(0.07) &    15.23(0.03) &  11.32(0.01)  & &   14.57(0.07) & 13.18(0.03)   &  11.32(0.02)   &&   \phn8.78(0.03)    &   \phn7.80(0.03)   &    \phn6.86(0.02)     &  5.77(0.04) \\
S5  \dotfill  &   IRSW-127    &  05:38:39.68  & --69:05:37.9  &  \nodata     & $>$19          &  13.39(0.05)  & &   \nodata     & \nodata       &  14.0\phn(0.03)    &&   \phn9.47(0.05)    &   \phn8.04(0.04)   &    \phn6.95(0.02)     &  6.0\phn(0.09)  \\
S6  \dotfill  &   P2          &  05:38:41.36  & --69:03:54.0  &  17.34(0.02) &    15.72(0.01) &  11.52(0.01)  & &   16.25(0.03) & 13.66(0.01)   &  11.5\phn(0.01)    &&   \phn9.13(0.03)    &   \phn8.39(0.03)   &    \phn7.65(0.02)     &  6.89(0.02) \\
S7A \dotfill  &   IRSN-122    &  05:38:46.84  & --69:05:05.4  &  19.50(0.25) &    16.83(0.08) &  13.13(0.02)  & &   16.13(0.09) & 14.81(0.05)   &  12.99(0.02)   &&   10.26(0.06)   &   \phn9.31(0.05)   &    \phn8.25(0.05)     &  \nodata    \\
S7B \dotfill  &   IRSN-126;P1 &  05:38:47.02  & --69:05:01.7  &  \nodata     &    17.59(0.04) &  11.97(0.01)  & &   15.59(0.1)\phn  & 13.75(0.04)   &  \nodata       &&   \phn8.81(0.04)    &   \phn7.84(0.03)   &    \phn7.01(0.03)     &  6.0\phn(0.06)  \\
S8  \dotfill  &   \dotfill    &  05:38:48.17  & --69:04:11.7  &  \nodata     &  \nodata       &  \nodata      & &   \nodata     & \nodata       &  \nodata       &&   10.11(0.06)   &   \phn9.33(0.06)   &    \phn7.8\phn(0.06)      &  6.01(0.07) \\
              &   IRSN-134    &  05:38:48.00  & --69:04:12.7  &  17.46(0.10) &    16.68(0.08) &  14.69(0.10)  & &   16.1\phn(0.1)\phn   & 15.64(0.12)   &  14.57(0.11)   &&   \nodata       &   \nodata      &    \nodata        &  \nodata    \\
              &   IRSN-137;P4 &  05:38:48.20  & --69:04:11.4  &  17.14(0.09) &    16.36(0.05) &  13.41(0.02)  & &   16.03(0.07) & 14.86(0.05)   &  13.28(0.04)   &&   \nodata       &   \nodata      &    \nodata        &  \nodata    \\
S9  \dotfill  &   IRSN-152    &  05:38:49.27  & --69:04:44.4  &  \nodata     & $>$19          &  14.51(0.05)  & &   \nodata     & 16.0\phn(0.09)    &  14.14(0.06)   &&   10.3\phn(0.07)    &   \phn9.12(0.06)   &    \phn7.55(0.08)     &  5.96(0.11) \\
S10 \dotfill  &   \dotfill    &  05:38:56.58  & --69:04:17.3  &  \nodata     &  \nodata       &  \nodata      & &   \nodata     & \nodata       &  \nodata       &&   11.59(0.09)   &   \nodata      &    \phn7.98(0.05)     &  6.86(0.10) \\
              &   S10A        &  05:38:56.31  & --69:04:16.1  &  19.9\phn(0.2)\phn   &    18.68(0.1)\phn  &  15.41(0.04)  & &   \nodata     & 16.9\phn(0.05)    &  14.99(0.03)   &&   \nodata       &   \nodata      &    \nodata        &  \nodata    \\
              &   S10B        &  05:38:56.60  & --69:04:17.5  &  \nodata     & $>$19          &  15.84(0.02)  & &   \nodata     & 18.64(0.26)   &  15.81(0.05)   &&   \nodata       &   \nodata      &    \nodata        &  \nodata    \\
              &   S10C        &  05:38:56.84  & --69:04:17.0  &  \nodata     & $>$19          &  15.31(0.03)  & &   \nodata     & 17.28(0.06)   &  15.23(0.03)   &&   \nodata       &   \nodata      &    \nodata        &  \nodata    \\
S10K \dotfill &               &  05:38:58.38  & --69:04:21.6  &  19.61(0.1)\phn  &    18.58(0.04) &  14.47(0.01)  & &   17.72(0.07) & 16.53(0.03)   &  14.55(0.02)   &&   11.82(0.05)   &   10.77(0.04)  &    \phn9.79(0.16)     &  \nodata    \\
S11  \dotfill &   \dotfill    &  05:38:52.67  & --69:04:37.5  &  \nodata     &  \nodata       &  \nodata      & &   \nodata     & \nodata       &  \nodata       &&   10.99(0.07)   &   10.5\phn(0.07)   &   \phn 8.6\phn(0.07)      &  6.98(0.08) \\
              &   IRSN-169    &  05:38:52.72  & --69:04:37.5  &  17.74(0.05) &    16.70(0.03) &  13.68(0.01)  & &   16.47(0.04) & 15.17(0.03)   &  13.84(0.02)   &&   \nodata       &   \nodata      &    \nodata        &  \nodata    \\
              &   IRSN-170    &  05:38:53.06  & --69:04:37.4  &  \nodata     &    18.45(0.08) &  15.41(0.05)  & &   18.12(0.1)\phn  & 16.73(0.07)   &  15.28(0.07)   &&   \nodata       &   \nodata      &    \nodata        &  \nodata    \\                         \enddata                                                                           \end{deluxetable}
\clearpage
\voffset0pt
\hoffset0pt

\begin{deluxetable}{ll|cc}
\tablenum{1b}
\tablecaption{Source Identifications in the {\it Spitzer} SAGE and IRSF Surveys}
\tabletypesize{\scriptsize}
\tablewidth{0pt}
\tablehead{
\colhead{{\it Spitzer}} &
\colhead{Alternate} &
\colhead{SAGE} &
\colhead{IRSF} \\
\colhead{Source} &
\colhead{ID's} &
\colhead{IRAC\tablenotemark{a}} &
\colhead{} }
\startdata
S1  \dotfill  &               & 053831.62-690214.6              &\nodata            \\
S2  \dotfill  &  IRSW-11      & SSTISAGEMA J053833.09-690611.7  & 05383309-6906118  \\
S3  \dotfill  &               & SSTISAGEMA J053834.05-690452.2  & 05383407-6904524  \\
S3K \dotfill  &               & SSTISAGEMA J053834.69-690450.0  & 05383475-6904504  \\
S4  \dotfill  &  IRSW-30;P3   & SSTISAGEMA J053834.60-690556.8  & 05383463-6905569  \\
S5  \dotfill  &  IRSW-127     & SSTISAGEMA J053839.68-690537.9  & 05383970-6905382  \\
S6  \dotfill  &  P2           & SSTISAGEMA J053841.36-690354.0  & 05384138-6903541  \\
S7A \dotfill  &  IRSN-122     & SSTISAGEMA J053846.84-690505.4  & 05384685-6905058  \\
S7B \dotfill  &  IRSN-126;P1  & SSTISAGEMA J053847.02-690501.7  & 05384701-6905019  \\
S8  \dotfill  &  \dotfill     & 053848.17-690411.7              & multiple          \\
              &  IRSN-134     & \nodata                         & 05384800-6904127  \\
              &  IRSN-137;P4  & \nodata                         & 05384820-6904114  \\
S9  \dotfill  &  IRSN-152     & 053849.27-690444.4              & 05384930-6904444  \\
S10 \dotfill  &  \dotfill     & SSTISAGEMA J053856.58-690417.3  & multiple          \\
              &  S10A         & \nodata                         & 05385631-6904161  \\
              &  S10B         & \nodata                         & 05385660-6904175  \\
              &  S10C         & \nodata                         & 05385684-6904170  \\
S10K \dotfill &               & SSTISAGEMA J053858.38-690421.6  & 05385837-6904214  \\
S11  \dotfill &  \dotfill     & 053852.67-690437.5              & multiple          \\
              &  IRSN-169     & \nodata                         & 05385272-6904375  \\
              &  IRSN-170     & \nodata                         & 05385306-6904374  \\                 
\enddata
\tablenotetext{a}{IRAC designations are from the SAGE IRAC SMP Archive 
(preceded by ``SSTISAGEMA J'') or from Gruendl \& Chu (2009).}
\end{deluxetable}
\clearpage

\section{Individual Sources}

Figures~4a, b, c display enlargements of the individual source fields, which 
show a number of significant local structures and relationships that are 
discussed in this section. 

\begin{figure}
\figurenum{4a}
\epsscale{0.9}
\includegraphics[width=\textwidth]{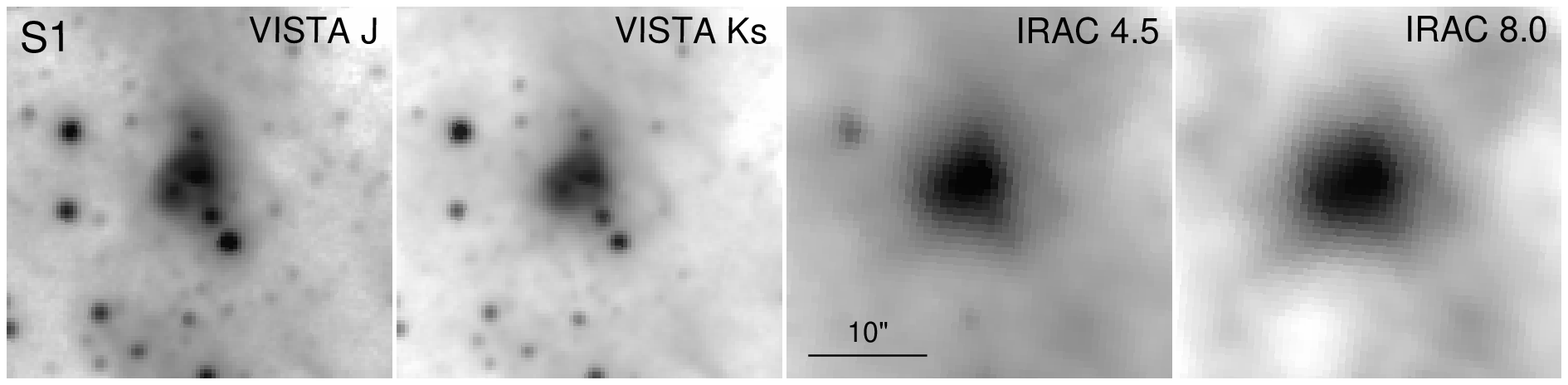}
\includegraphics[width=\textwidth]{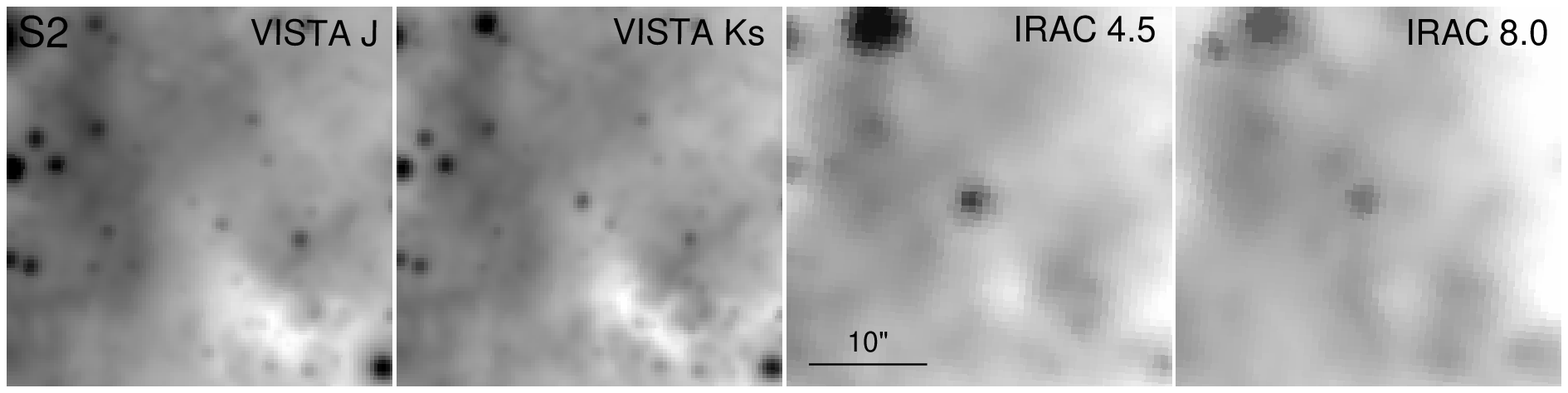}
\includegraphics[width=\textwidth]{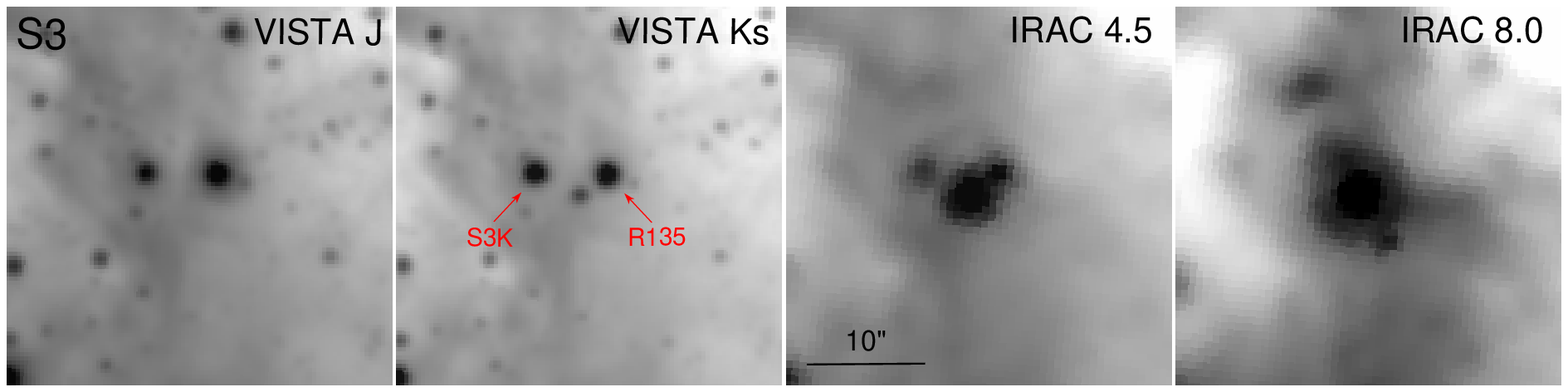}
\includegraphics[width=\textwidth]{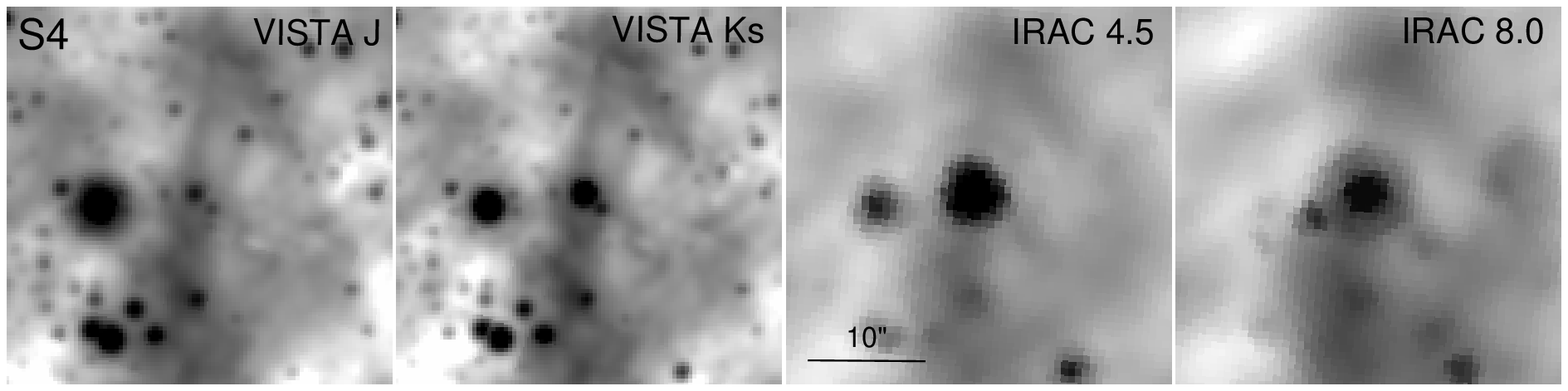}
\includegraphics[width=\textwidth]{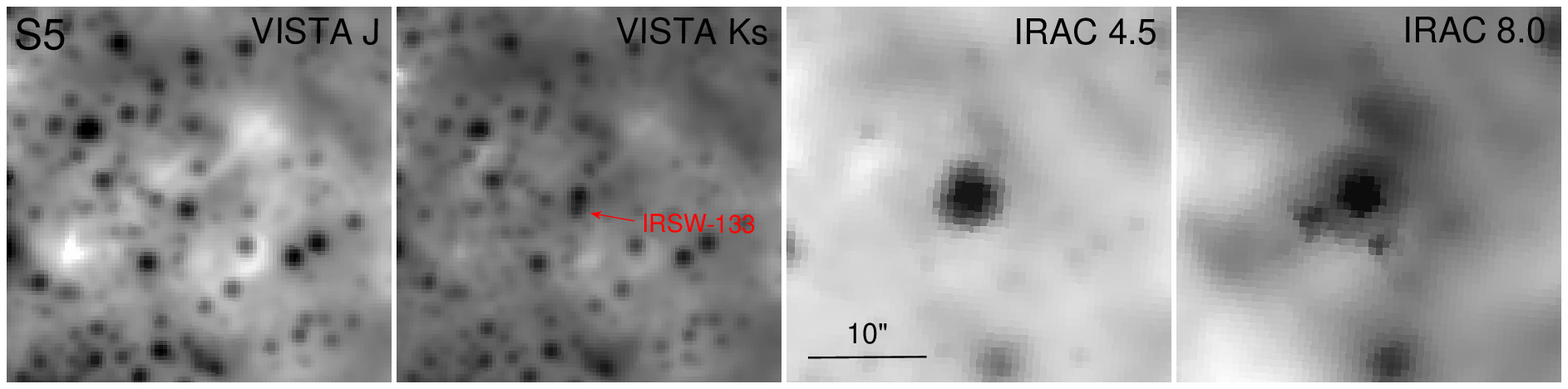}
\caption{Larger scale cutouts of the present IR sources in two VISTA/VMC and two {\it Spitzer\/}/IRAC bands, which reveal additional structural details.  
North is up and east to the left in each panel.  Among other things, note here 
the reversal of relative magnitudes in the S3 system between the two wavelength regimes, and the invisibility of S2, S3, and S5 in $J$.  R135 is a luminous WN star.}
\end{figure}

\begin{figure}[t]
\figurenum{4b}
\epsscale{0.9}
\includegraphics[width=\textwidth]{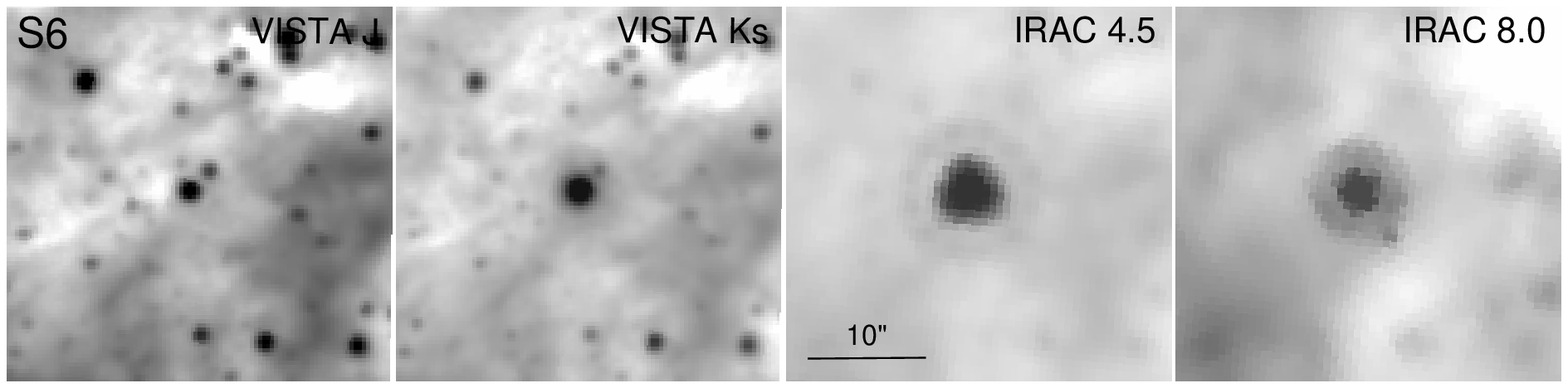}
\includegraphics[width=\textwidth]{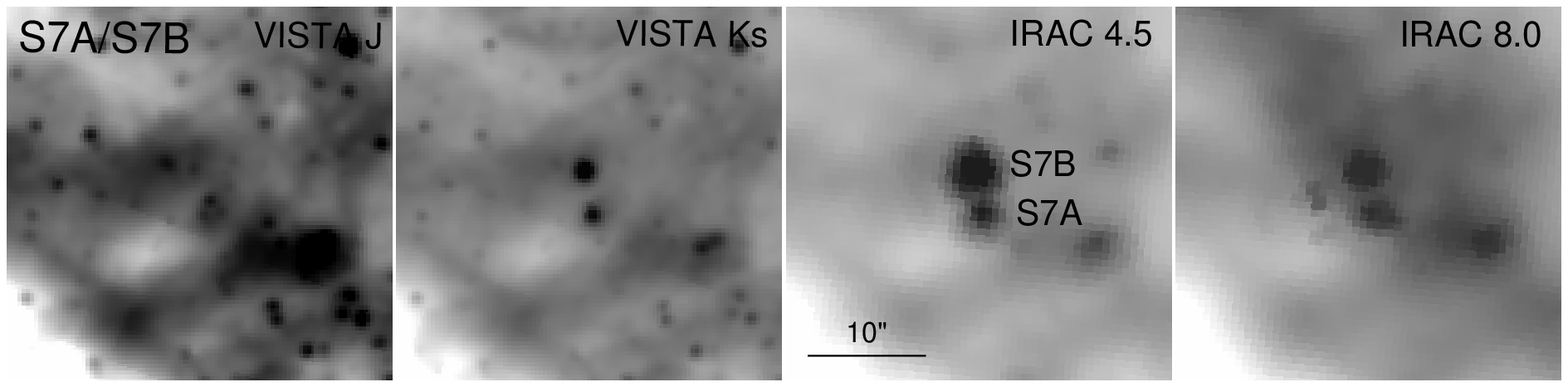}
\includegraphics[width=\textwidth]{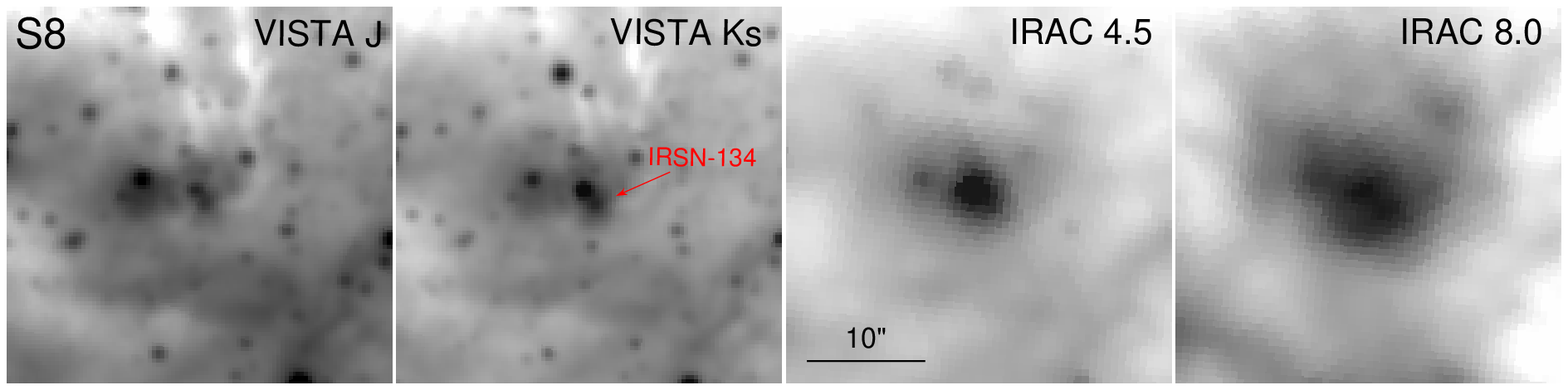}
\caption{As Fig.~4a.  The object 2\arcsec\ NW of S6 is an LPV that is brighter 
than the YSO at shorter wavelengths, except when in dust-formation episodes.  
See {\it HST\/} images of S8 in Walborn et~al.\ (2002) that show the associated 
cluster of faint IR sources.}
\end{figure}

\begin{figure}[t]
\figurenum{4c}
\epsscale{0.9}
\includegraphics[width=\textwidth]{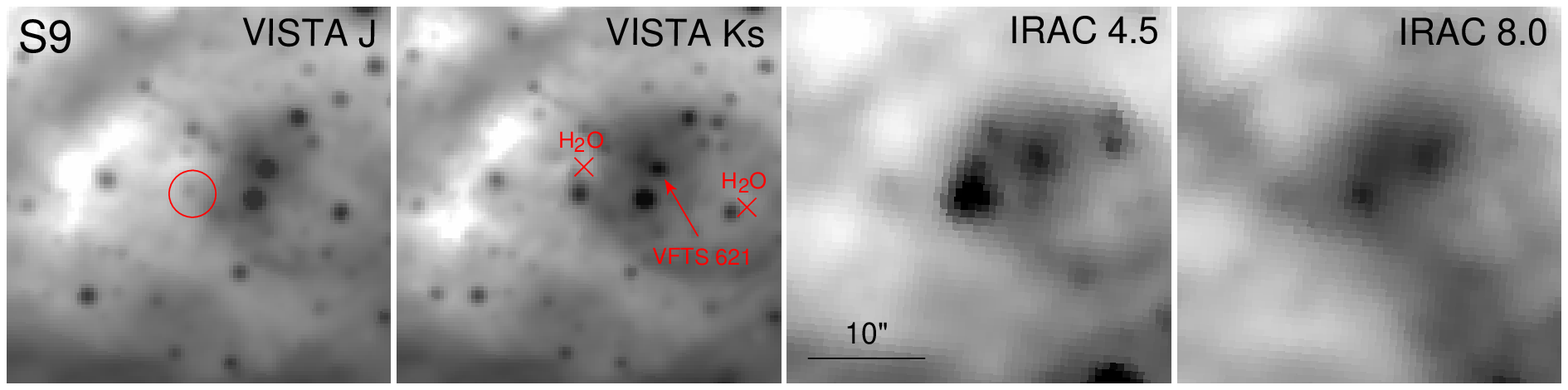}
\includegraphics[width=\textwidth]{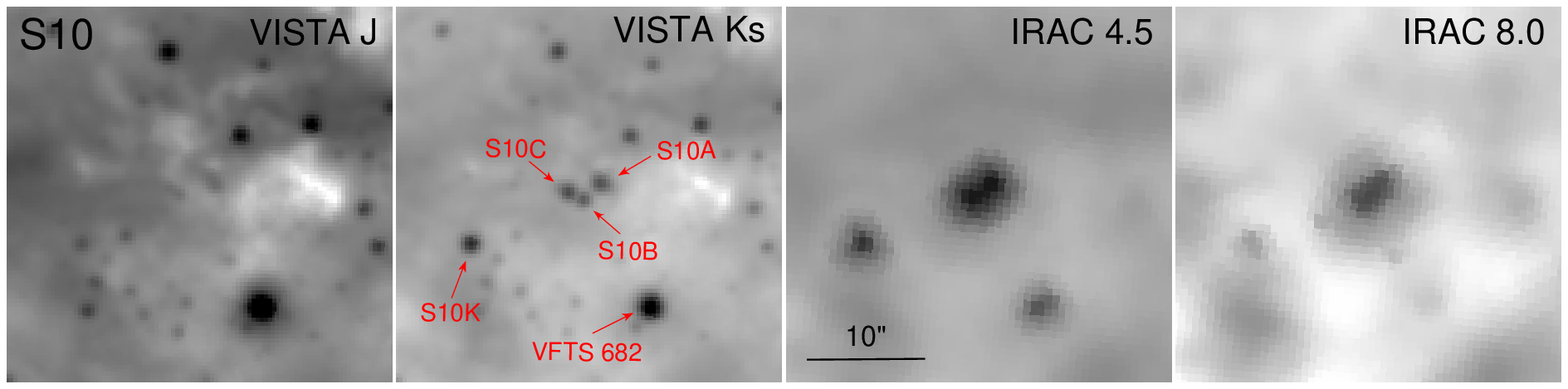}
\includegraphics[width=\textwidth]{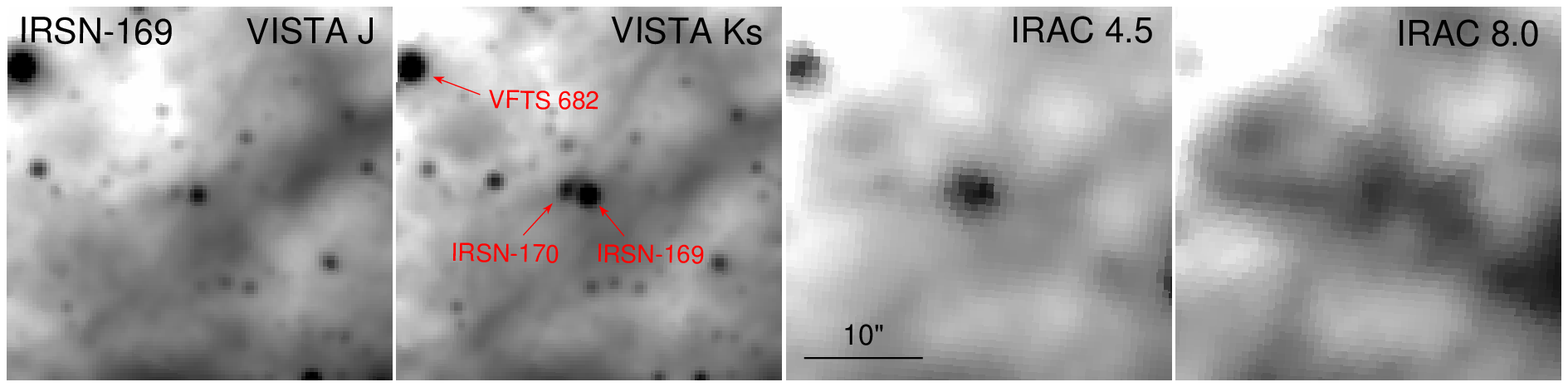}
\caption{As Fig.~4a.  See Walborn et~al.\ (1999) for the remarkable extended 
structure of S9 in {\it HST\/}/NICMOS images; VFTS 621 is a very young O2 star.  The positions of two water masers reported by Ellingsen et~al.\ (2010) in
this field are also marked.  VFTS~682 is a luminous WN star.  Rubio et~al.\ (1998) IRSN-169 is discussed with S10 in Section~3.10.}
\end{figure}

\subsection{S1} \label{}

We were surprised to find one of the most luminous {\it Spitzer\/} sources
embedded in extensive diffuse emission at the northwestern extreme of the 
field, evidently unrelated to R136.  S1 appears to be associated with a larger CO feature catalogued as 30Dor-06 by Johansson et~al.\ (1998), and with an H~II region listed as No.~889 by Kastner et~al.\ (2008).  In the IR images it appears as a somewhat diffuse double source, with the fainter barely resolved 
component toward the SE.  Fortunately, a recent {\it HST\/}/WFC3 F775W image (GO~12499, PI D.J.~Lennon) reveals what lurks within (Figure~5): the object appears as a dusty H~II knot, here named ``The Skull Nebula,'' containing a compact cluster dominated by four $I$ $\sim$ 19--21 stars in an EW line (the brighter IR ``component'') and two more toward the SE (the fainter IR ``component'') (E.~Sabbi, priv.\ comm.).  Two more comparable stars within the outer nebulosity to the N and S may be additional members or field objects.     

\begin{figure}[t]
\figurenum{5}
\epsscale{1.0}
\plotone{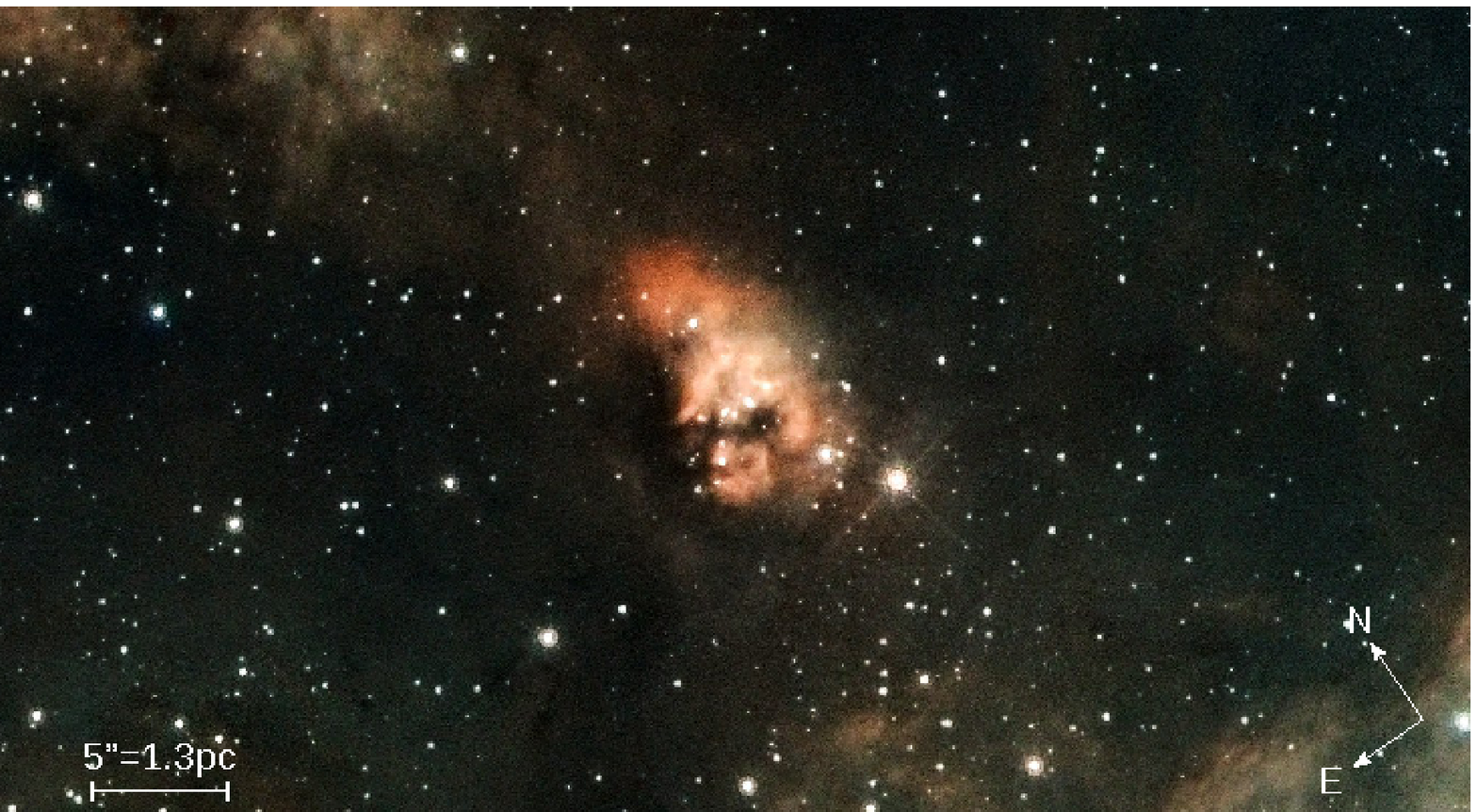}
\caption{\label{fig:fig5} S1, ``The Skull Nebula,'' as seen in the 
{\it HST\/}/WFC3/F775W + groundbased composite image presented by E.~Sabbi in STScI news release 2012-35 
(http://hubblesite.org/\break newscenter/archive/releases/2012/35/).  
See text for discussion.}
\end{figure}

With reference to Figures~1 and 2, we propose an origin of star formation
at this location triggered by adjacent SN events.  To the NE of S1 are the 
three luminous, ``isolated'' WN stars R144, R146, and R147 (Feast et~al.\ 
1960), apparently associated with the large kinematical (Chu \& Kennicutt 
1994) and X-ray (Townsley et~al.\ 2006) shell at the north of 30~Dor.  This 
object could have been produced by the explosion of another former WR star.  
To the SW of S1 lies the 20~Myr old cluster Hodge~301 that has produced 40~SN (Grebel \& Chu 2000) and is surrounded by another giant shell.  Numerous 
dark globules can also be seen along the interface between the two shells 
in optical images, suggesting that further star formation is likely to occur 
in these compressed filaments---a novel feature of 30~Doradus.  

\subsection{S2} \label{}

This relatively fainter source is ISRW-11 of Rubio et~al.\ (1998).  (It is 
also NIC07a of Brandner et~al.\ 2001, in which the cross-identification with 
Rubio et~al.\ was missed.)  It is located at the northern end of a diffuse 
structure that is bright in the {\it Spitzer\/} image (Fig.~1) but dark in \textit{JHK} (Fig.~2).  As noted by Walborn et~al.\ (1999), with appropriate stretch and scale the object appears multiple in the NICMOS data, and analysis of its PSF indicates that it is extended, with a Gaussian sigma of 1.29~pix vs.\ 1.04~pix for a point source in~$K$.  Here we analyze the S2 photometry as if from a 
single source, with the caveat that it may actually be a tight multiple system.

\subsection{S3} \label{}

This luminous source apparently within a Tarantula filament is at the apex 
of an inverted triangular asterism with the ``isolated'' WN star R135 (Feast 
et~al.\ 1960) and another source brighter at shorter wavelengths that we 
designate S3K.  The reversal of their relative magnitudes between Figures~1 
and 2 is striking; see also Figure~4a.  We suggest that the massive wind of 
the WN star may have triggered the formation of S3 in the adjacent nebular 
filament.  As further analyzed below, S3K may be a star rather than a YSO.

\subsection{S4} \label{}

S4 is one of the most luminous sources in 30~Dor at all NIR wavelengths, also 
IRSW-30 of Rubio et~al.\ (1998), NIC03a of Brandner et~al.\ (2001), and P3 of  Hyland et~al.\ (1992).  As extensively discussed based on {\it HST\/} optical 
and IR images by Walborn et~al.\ (1999, 2002), it is located in the head of a 
large, bright-rimmed dust pillar oriented toward R136.  They measured magnitude 20.3 in $V$ and 19.3 in $I$; here we report 11.3 in $K$ and 5.8 at 8~$\micron$ (Table~1a).  This is a prime object for astrophysical investigation, including spectroscopy from visual through IR.  IRSW-26 is a companion to the SW, but it is more than 3~mag fainter in $K$ and evidently not significant in the {\it Spitzer\/} bands (Figure~4a), so we analyze S4 as a single object.

\subsection{S5} \label{}

S5 or Rubio et al. (1998) IRSW-127 is located within a prominent boomerang- (or \hbox{stapler-)} shaped dust cloud silhouetted against the bright stellar and nebular field just north of R136 in optical images.  Again, this cloud is bright in the {\it Spitzer\/} image.  It is likely located in the immediate foreground of R136. The source IRSW-133 immediately to the south is brighter at $J$, in which S5 is invisible (Figure~4a), but the latter is brighter at $K$ and likely dominates in the {\it Spitzer\/} images.  These sources were  also measured at $H$ and $K_{\rm s}$ by Campbell et~al.\ (2010, their Section 5.4), with results similar to ours, although the three available $K$~magnitudes are in relatively poor agreement, perhaps due to source confusion or variability.  C.~Evans (priv.\ comm.) has pointed out that IRSW-133,  which Campbell et~al.\ suggested to be a massive star, is VFTS~476 in the current VLT-FLAMES Tarantula Survey (Evans et~al.\ 2011); indeed, it has been classified as O((n)) from the presence of He~II absorption lines in the VFTS data (Walborn et~al.\ in prep.), with heavy nebular emission-line contamination preventing a more precise description.  Seale et~al.\ (2009) report {\it Spitzer\/} spectroscopy of this source consistent with a YSO nature.

\subsection{S6} \label{}

S6 is the very luminous P2 of Hyland et~al.\ (1992) and NIC16a of Brandner 
et~al.\ (2001).  It is a point source in {\it HST\/} optical and IR images, as 
shown by Walborn et~al.\ (2002) who measured an $I$~mag of 20.9 and placed a $V$~limit of 23.  Beware of potential confusion with a Long Period Variable (LPV) about 2\arcsec\ NW, which oscillates around 17th mag in $I$ except during dust-formation episodes when it becomes fainter than S6 (OGLE-LMC-LPV-76683; Soszynski et~al.\ 2009).  The LPV is always much fainter than S6 at longer wavelengths (Figure~4b). To within the {\it HST\/} resolution, S6 appears to be an instance of monolithic, ``isolated'' massive star formation.

\subsection{S7} \label{}

This intense ``double source'' is Rubio et~al.\ (1998) IRSN-122/126, the latter dominating Hyland et~al.\ (1992) P1; it is also Brandner et~al.\ (2001) NIC12b/d.  As extensively discussed by Walborn et~al.\ (1999, 2002), they are embedded within a massive dust pillar oriented toward R136 and parallel to the adjacent Knot~1 pillar from which a young stellar system has just emerged.  Like S4 and S6, these luminous sources can be measured at $I$ (20.4 for S7A and 21.2 for S7B), with $V$~limits of 22 for both; hence they are also outstanding candidates for future spectroscopic investigation over a wide wavelength range.

\subsection{S8} \label{}

In striking contrast to S6, this bright source (P4 of Hyland et~al.\ 1992; IRSN-137 of Rubio et~al.\ 1998; NIC15b of Brandner et~al.\ 2001, but
mislabeled there as IRSN-134) is surrounded by a rich cluster of much fainter ones as shown by Walborn et~al.\ (2002), indicating qualitatively different modes of massive star formation for these two sources.  This circumstance creates difficulties with the photometry of S8, because the VISTA and IRSF images from which the \textit{YJHK} magnitudes listed in Table~1a are derived resolve it from the cluster, whereas the {\it Spitzer\/} images do not.  Rubio et~al.\ distinguish a relatively bright ``tail'' of the blended cluster toward the SW as the separate source IRSN-134, but only the NICMOS images reveal the actual structure of the cluster.  We have noted that the cluster is surrounded by a thick annulus of emission in the {\it Spitzer\/}/MIPS 24~$\micron$ image, which warrants further study.

\subsection{S9} \label{}

This remarkable source is a constituent of the intricate YSO/ZAMS phenomena associated with the Knot~2 complex of Walborn \& Blades (1997).  With reference to Figure~3 and Section~3.2 of Walborn et~al.\ (1999): S9 is the brighter of two extended sources aligned across and approximately equidistant from the central, young optical stellar system (Walborn et~al.\ 2002).  The major axis of S9 has an extent of $1\arcsec$, or 0.25~pc~$\sim$~50,000~AU. Both extended sources display (opposite) color gradients perpendicular to their radius vectors from the central system, suggesting (counterclockwise) rotation.  No evidence of the presumed jets impacting the surrounding dust clouds to produce them has been found to date.  The spectral type of the brightest stellar component in Knot~2, VFTS~621 or Parker (1993) 1429, has been found to be O2~V((f*))z from the VFTS data (Walborn et~al.\ in prep.), thus a very massive young object.  

It is surprising that S9 would be one of the brightest {\it Spitzer\/} sources in 30~Dor, but then, its nature is entirely unknown.  Seale et~al.\ (2009) report {\it Spitzer\/} spectroscopy of this source consistent with a YSO, although the slit width is at least $\sim$$4\arcsec$ and S9 is not dominant at 8~$\micron$ (Fig.~4c). Again, the NICMOS images show that it is nonstellar, so that a very large circumstellar envelope would be implied in that case, but the point-symmetric counterpart on the opposite side of the central stellar system suggests a more exotic interpretation.  Moreover, we have found that S9 and Knot~2 are apparently associated with the brightest 30~Dor source in all of the {\it Herschel\/} 100--500~$\micron$ bands, from the HERschel Inventory of The Agents of Galaxy Evolution (HERITAGE; Meixner et~al.\ 2010).  Finally, there are also two water masers associated with this complex (Ellingsen et~al.\ 2010), one of which is very near S9, while the other is roughly opposite across the central stellar system, approximately mimicking the alignment of the IR ``mystery spots'' at a lower PA (Figure~4c; Walborn et~al.\ 1999).  Clearly further detailed, combined analysis of the multiple rich datasets now available for this region will prove rewarding.   

\subsection{S10} \label{}

At first glance S10 appears to be one of the brightest sources in the {\it Spitzer\/} images, but on closer inspection it is somewhat extended, and indeed, the higher resolution VISTA images reveal it as three much fainter sources that we designate A, B, C.  Moreover, they may comprise part of a larger structure associated with the luminous WN star VFTS~682, newly discovered by that survey (Bestenlehner et~al.\ 2011). The WN star appears to have excavated a cavity, around the periphery of which there is an array of IR sources, including the S10 system.  Another bright $K$~source that we designate S10K forms a triangular asterism with VFTS~682, and S10 at the apex (Figure~4c).  Rubio et~al.\ (1998) IRSN-169 and its fainter NE companion IRSN-170 lie near the opposite side of the cavity (midway between VFTS~682 and S9 in Figures~1 and 2); the blended {\it Spitzer\/} data for the two sources are ascribed to ``S11'' in Table~1 and Fig.~3.  There is also a water maser exactly coincident with IRSN-169 (Ellingsen et~al.\ 2010).  Thus, along with S3/R135, this structure may represent another case of a massive WN wind triggering further star formation. 

\section{Discussion}

\subsection{Model Fitting and Masses} \label{}

We fitted YSO models to the SEDs of selected, unresolved sources. We used the Robitaille et~al.\ (2006) grid of 20,000 2D YSO radiative transfer models for masses from 0.1 through 50~$M_{sun}$.  For each model in the grid, SEDs are calculated at 10 viewing angles, resulting in 200,000 pre-computed SEDs that can be compared to the observations using the Robitaille et~al.\ (2007) fitting tool.  Since these models are computed for single objects, we selected those sources from our sample that appear single in the near-IR and mid-IR images: S2, S3, S3K, S4, S6, S7A, S7B, and S10K.  For the fitting, we used VISTA~$J$ and $K_{\rm s}$, IRSF $H$, and IRAC 3.6--8.0 $\mu$m fluxes as available (Table~1a), as well as {\it HST\/}/WFPC2 F814W ($\sim$$I$) and F555W ($\sim$$V$) fluxes or upper limits for S4, S6, S7A, and S7B (these magnitudes are cited in the individual source discussions above).  The fits for these sources are shown in Figure~6.  All of the YSO fits indicate Class~I objects.

The formal fit for S3K as a star is superior to that as a YSO, so both fits are shown for this source.  This result is related to the fact that the {\it Spitzer\/} fluxes are smaller than those at shorter wavelengths, unlike all the other cases.  The stellar fit yields $T_{\rm eff}$~=~4750~K and A$_V=5.85$.  While the apparent close association of S3K with S3, R135, and nebular structure shown in the figures may be a chance alignment, there is also the interesting possibility that it could be a post-YSO, PMS object.

\begin{figure}[t]
\figurenum{6}
\epsscale{1}
\plotone{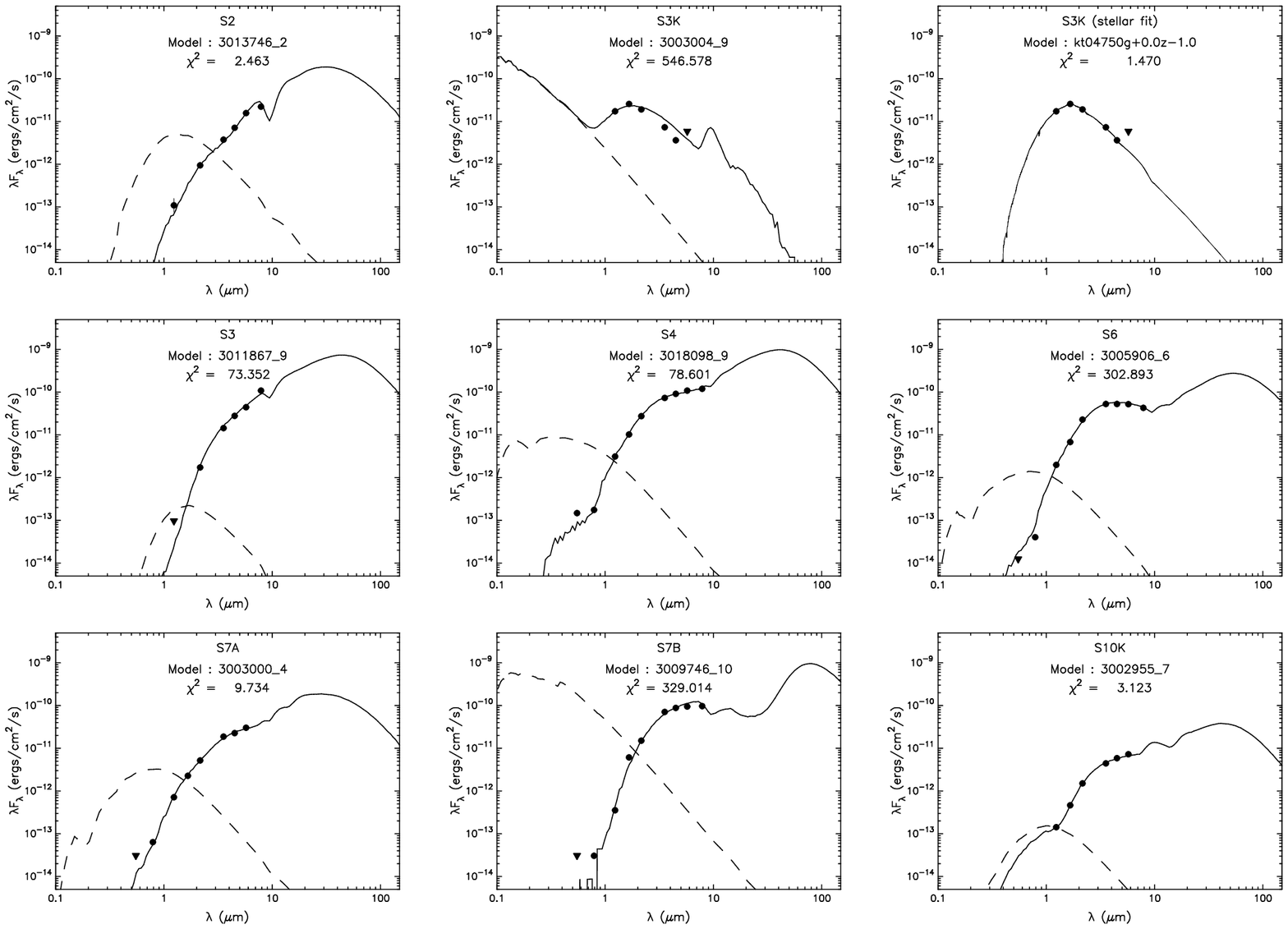}
\caption{\label{fig:fig6} SEDs of eight apparently single IR sources.  The solid black line shows the best-fitting model.  The dashed line corresponds to the (extincted) SED of the stellar photosphere in the best-fitting model.  S3K has been fitted as both a YSO and a star, with the latter fit appearing preferable.  The plots are labeled with Robitaille et~al.\ (2006) model numbers and global $\chi^2$ values.  Further parameters resulting from the fits are given in Table~2.}
\end{figure}

The physical parameters estimated from the fitting are listed in Table~2: best and average stellar luminosity, effective temperature, and mass; envelope and disk masses; and interstellar extinction. Since the Robitaille et~al.\ (2006) models assume a Galactic gas-to-dust ratio, we scaled the results for the envelope and disk masses to the 30~Dor ratio by multiplying them by a factor of 2.1 (Dobashi et~al.\ 2008).  The averages are over all YSO model fits with $\chi^{2}$ per data point between $\chi^{2}_\mathrm{best}$ and $\chi^{2}_\mathrm{best}$~+~2, as also listed in the table.  The uncertainties of the averages quoted (in parentheses) in Table~2 represent standard deviations of the means.  For sources with very high $\chi^{2}$/pt and only one fitted model (S6 and S7B), there are of course no averages and the derived parameters are highly uncertain.  For other fits with $\chi^{2}$/pt~$>$~5, the physical parameters may not reflect well the actual properties of the sources.  Source S3K is also listed in Table~2 as a YSO and only its luminosity is provided; however, as already noted this source is more likely to be a star with the parameters given above.  The $\chi^{2}_\mathrm{min}$/pt for the stellar photosphere fit is $\sim$6, far lower than that for the YSO fit, $\sim$109 as listed in the table. 

The models assume single sources; at the distance of the LMC, even the sources that appear unresolved in our data could in fact be multiple or even protoclusters (see also Carlson et~al.\ 2011). Our derived parameters are valid only on the assumption that a single source dominates the luminosity at all wavelengths.  

\clearpage
\thispagestyle{empty}
\voffset0.3in
\hoffset1in
\begin{deluxetable}{p{1.2cm}ccp{0.1cm}ccccccccccccccccc}
\tabletypesize{\scriptsize}
\rotate
\tablewidth{0pt}
\tablenum{2}
\tablecaption{The SED Fitting Results: Physical Parameters}

\tablehead{
\colhead{Source} &
\multicolumn{2}{c}{L$_{\star}$ (10$^{4}$ L$_{\odot}$)} &
\colhead{} &
\multicolumn{2}{c}{T$_{\star}$ (10$^{4}$ K)} &
\colhead{} &
\multicolumn{2}{c}{M$_{\star}$ (M$_{\odot}$)} &
\colhead{} &
\multicolumn{2}{c}{M$_\mathrm{env}$ (10$^{2}$ M$_{\odot}$)} &
\colhead{} &
\multicolumn{2}{c}{M$_\mathrm{disk}$ (M$_{\odot}$)} &
\colhead{} &
\multicolumn{2}{c}{A$_{V}$} &
\colhead{} &
\colhead{$\chi_{\rm min}^{2}$/pt\tablenotemark{b}} &
\colhead{n$_{\rm fits}$\tablenotemark{c}} \\

\cline{2-3}
\cline{5-6}
\cline{8-9}
\cline{11-12}
\cline{14-15}
\cline{17-18}

\colhead{Name} &
\colhead{best} &
\colhead{ave} &
\colhead{} &
\colhead{best} &
\colhead{ave} &
\colhead{} &
\colhead{best} &
\colhead{ave} &
\colhead{} &
\colhead{best} &
\colhead{ave} &
\colhead{} &
\colhead{best} &
\colhead{ave} &
\colhead{} &
\colhead{best} &
\colhead{ave} &
\colhead{} &
\colhead{} &
\colhead{} 
}
\startdata
      S2 \dotfill  &      3.4  &    7.7(0.7)\rlap{\tablenotemark{a}}  &&    1.2  &    2.1(0.1)  &&    20.1  &   23.0(0.7)  &&    \phn2.2  &    6.0(0.7)  &&   0     &   0.15(0.05)  &&    6.5  &    5.4(2.7)  &&    \phn0.4  &    \llap{10}9  \\
      S3 \dotfill  &      8.2  &   \llap{1}0.5(2.5)  &&    3.8  &    3.4(0.6)  &&    25.2  &   26.7(3.2)  &&   13.3  &   \llap{1}4.4(7.7)  &&   0     &   0           &&   \llap{1}0.0  &    9.7(0.4)  &&   14.7  &       3  \\
     S3K \dotfill  &      1.7  &    \nodata   && \nodata &    \nodata   &&  \nodata &   \nodata    && \nodata &   \nodata    &&\nodata  &   \nodata     &&\nodata  &    \nodata   &&  \llap{1}09.3  &       1  \\
      S4 \dotfill  &     10.7  &    8.2(2.5)  &&    3.9  &    3.8(0.1)  &&    27.4  &   24.5(2.9)  &&   10.9  &    6.4(4.5)  &&   0.19  &   0.10(0.09)  &&    1.8  &    1.2(0.6)  &&    \phn8.7  &       2  \\
      S6 \dotfill  &      3.7  &    \nodata   &&    3.4  &    \nodata   &&    18.4  &   \nodata    &&   12.0  &   \nodata    &&   0.13  &   \nodata     &&    3.0  &    \nodata   &&   37.9  &       1  \\
     S7A \dotfill  &      3.0  &    2.7(0.4)  &&    2.0  &    2.4(0.4)  &&    15.2  &   16.4(0.6)  &&    \phn1.6  &    3.1(1.2)  &&   0.04  &   0.13(0.09)  &&    3.2  &    2.8(1.7)  &&    \phn1.4  &       9  \\
     S7B \dotfill  &      7.3  &    \nodata   &&    1.7  &    \nodata   &&    23.4  &   \nodata    &&   71.5  &   \nodata    &&   0     &   \nodata     &&    0.1  &    \nodata   &&   41.1  &       1  \\
    S10K \dotfill  &      0.7  &    2.4(1.8)  &&    2.5  &    2.0(0.1)  &&     \phn9.1  &   10.9(1.5)  &&    \phn0.5  &    2.8(1.9)  &&   0.28  &   0.11(0.02)  &&    4.4  &    4.5(2.4)  &&    \phn0.5  &     \llap{2}4  \\
\enddata
\tablenotetext{a}{~The uncertainties represent standard deviations of the means.}
\tablenotetext{b}{~$\chi^{2}_{\rm min}\,$/pt  is a $\chi^{2}$ per data point for the best-fit model.}
\tablenotetext{c}{~n$_{\rm fits}$ is the number of fits with $\chi^{2}$/pt between $\chi^{2}_{\rm min}\,$/pt and $\chi^{2}_{\rm min}\,$/pt$+$2.}
\end{deluxetable}
\clearpage
\voffset0pt
\hoffset0pt

\section{Summary/Outlook}

Based primarily upon {\it Spitzer\/}/SAGE, VISTA/VMC, and IRSF data, with reference to {\it HST\/}/\break NICMOS/WFPC2/WFC3 images for higher spatial resolution, we have presented the first comprehensive description of the brightest IR point sources throughout the triggered star formation within the 30~Doradus Nebula in the LMC.  The {\it Spitzer\/} data extend prior work to longer wavelengths, and all of the subsequent IR images cover fields not included in the NICMOS dataset, thus completing the global IR view of 30~Dor. The overall spatial distribution of these sources is consistent with the two-stage starburst picture discussed previously, but the intricacy and diversity among the individual ``point'' sources revealed here is striking.  There are clear implications for all YSO studies in the Magellanic Clouds, not to mention at greater distances.  In summary, the most outstanding individual source results are listed below:
\begin{itemize}
\item S1 is a newly resolved compact H~II region containing  a multiple system of massive stars readily seen in the $I$~band, possibly triggered by adjacent SN events.  The object is named ``The Skull Nebula'' from its visual appearance in the WFC3 image.

\item S3, the multiple S10, and other nearby sources are apparently associated with isolated, luminous WN stars and may be the result of local triggering by the massive winds of the latter.   A possible analogous instance in the Galaxy was recently reported by Liu et~al.\ (2012).

\item S4, S7A, and S7B are located within massive dust pillars oriented toward R136, as discussed in previous work.

\item S6 is a very luminous, unresolved point source in the {\it HST\/} images and may represent an instance of isolated massive star formation.

\item S8 is the most luminous source in a small cluster of fainter objects, in marked contrast to S6.  The cluster is surrounded by a 24~$\micron$ shell in the {\it Spitzer\/}/MIPS image.

\item S9 is an unprecedented extended source of unknown nature; with its fainter counterpart on the opposite side of a very young, massive optical system, its morphological appearance suggests the impact point of an invisible, rotating jet on surrounding dust clouds.  Surprisingly, it is one of the brightest {\it Spitzer\/} sources, and it is related to the brightest {\it Herschel\/} source in the region.  There are also two water masers among the array of interesting objects in this configuration, which will likely provide unique information about massive star formation.   
\end{itemize}

We have fitted YSO models to the data for eight apparently (predominantly) 
single objects to derive physical parameters including stellar masses, which 
fall in the 10--30~$M_{\sun}$ range, in general agreement with the results of 
Hyland et~al.\ (1992) for four of the brightest sources.

These results can be extended in several ways, at least two of which are obvious.  First, measurements and analysis of the complete IR sample in this 
region should be carried out in order to estimate the total mass of the second, 
triggered generation in 30~Doradus, for comparison with the first generation 
to illuminate the general structure and evolution of starbursts on this scale.  

Second, broad-wavelength spectroscopy of the bright sources is essential to 
advance understanding of their diverse natures.  A current ESO/VLT/X-Shooter program (PI L.~Kaper) is addressing this objective.  As noted in the individual discussions, some of these sources are detectable at $I$ and even at $V$, which promises significant further insights into their physical 
properties.  For example, a comparable observation of a Galactic YSO has recently yielded photospheric absorption as well as circumstellar emission features from the optical through the NIR (Kaper et~al.\ 2011; Ochsendorf et~al.\ 2011).

\acknowledgments

This paper is based largely on (SAGE) observations made with the {\it Spitzer 
Space Telescope\/}, which is operated by the Jet Propulsion Laboratory, 
California Institute of Technology, under contract with NASA; and in part on 
(VISTA/VMC) data products from observations made with ESO Telescopes at the La~Silla or Paranal Observatories under ESO program ID~179.B-2003; and in part on data from the IRSF Magellanic Clouds Point Source Catalogue obtained at the SAAO.  Ancillary use has been made of data products from 2MASS, which is a joint project of the University of Massachusetts and the Infrared Processing and Analysis Center/California Institute of Technology, funded by NASA and NSF.  Thanks to Elena Sabbi and Danny Lennon for their unpublished {\it HST\/}/WFC3 data and discussion regarding S1.  M.S.\ acknowledges financial support from the NASA ADAP award NNX11AG50G.  Publication support was provided by NASA through grant GO-12465.01 (PI P.~Crowther) from STScI.


\begin{thebibliography}{}

\bibitem[]{} Bestenlehner, J.M. et al. 2011, \aap, 530, L14

\bibitem[]{} Brandner, W., Grebel, E.K., Barb\'a, R.H., Walborn, N.R., \&
Moneti, A. 2001, \aj, 122, 858

\bibitem[]{} Campbell, M.A., Evans, C.J., Mackey, A.D. et al. 2010, \mnras, 405, 421

\bibitem[]{} Carlson, L.R. et al. 2011, \apj, 730, 78

\bibitem[]{} Chu, Y.-H., \& Kennicutt, R.C., Jr. 1994, \apj, 425, 720

\bibitem[]{} Cioni, M.-R.L. et al. 2011, \aap, 527, A116 

\bibitem[]{} Dalton, G.B. et al. 2006, SPIE Conf.\ Ser., 6269, 30

\bibitem[]{} Dobashi, K., Bernard, J.-P., Hughes, A., Paradis, D., Reach, W.T., \& Kawamura, A. 2008, \aap, 484, 205

\bibitem[]{} Ellingsen, S.P., Breen, S.L., Caswell, J.L., Quinn, L.J., \& Fuller, G.A. 2010, \mnras, 404, 779 

\bibitem[]{} Emerson, J.P. et al. 2004, Proc.\ SPIE, 5493, 401

\bibitem[]{} Evans, C.J. et al. 2011, \aap, 530, A108

\bibitem[]{} Fazio, G.G. et al. 2004, \apjs, 154, 10

\bibitem[]{} Feast, M.W., Thackeray, A.D., \& Wesselink, A.J. 1960, \mnras, 121, 337

\bibitem[]{} Grebel, E.K., \& Chu, Y.-H. 2000, \aj, 119, 787

\bibitem[]{} Gruendl, R.A., \& Chu, Y.-H. 2009, \apjs, 184, 172

\bibitem[]{} Hyland, A.R., Straw, S., Jones, T.J., \& Gatley, I. 1992, \mnras, 257, 391

\bibitem[]{} Indebetouw, R. et al. 2009, \apj, 694, 84

\bibitem[]{} Irwin, M.J. et al. 2004, Proc. SPIE, 5493, 411

\bibitem[]{} Johansson, L.E.B. et al. 1998, \aap, 331, 857

\bibitem[]{} Kaper, L., Ellerbroek, L.E., Ochsendorf, B.B., \& Caballero Pouroutidou, R.N. 2011, AN, 332, 232

\bibitem[]{} Kato, D. et al. 2007, PASJ, 59, 615

\bibitem[]{} Kastner, J.H. et al. 2008, \aj, 136, 1221

\bibitem[]{} Liu, T., Wu, Y., Zhang, H., \& Qin, S.-L 2012, \apj, 751, 68

\bibitem[]{} Meixner, M. et al. 2006, \aj, 132, 2268

\bibitem[]{} Meixner, M. et al. 2010, \aap, 518, L71

\bibitem[]{} Ochsendorf, B.B., Ellerbroek, L.E., Chini, R., Hartoog, O.E., Hoffmeister, V., Waters, L.B.F.M., \& Kaper, L. 2011, \aap, 536, L1

\bibitem[]{} Oliveira, J.M. et al. 2009, \apj, 707, 1269

\bibitem[]{} Parker, J.Wm. 1993, \aj, 106, 560

\bibitem[]{} Rieke, G.H. et al. 2004, \apjs, 154, 25

\bibitem[]{} Robitaille, T.P., Whitney, B.A., Indebetouw, R., Wood, K., \& Denzmore, P. 2006, \apjs, 167, 256

\bibitem[]{} Robitaille, T.P., Whitney, B.A., Indebetouw, R., \& Wood, K. 2007, \apjs, 169, 328   

\bibitem[]{} Rubio, M., Barb\'a, R.H., Walborn, N.R., Probst, R.G., Garc\'{\i}a, J., \& Roth, M.R. 1998, \aj, 116, 1708 

\bibitem[]{} Rubio, M., Roth, M., \& Garc\'{\i}a, J. 1992, \aap, 261, L29

\bibitem[]{} Scowen, P.A. et al. 1998, \aj, 116, 163

\bibitem[]{} Seale, J.P. et al. 2009, \apj, 699, 150

\bibitem[]{} Shimonishi, T. et al. 2008, \apj, 686, L99

\bibitem[]{} Skrutskie, M.F. et al. 2006, \aj, 131, 1163 

\bibitem[]{} Soszynski, I. et al. 2009, Acta Astron, 59, 239

\bibitem[]{} Stetson, P.B. 1987, \pasp, 99, 191

\bibitem[]{} Townsley, L.K. et al. 2006, \aj, 131, 2140 

\bibitem[]{} Walborn, N.R. 2002, in ASP Conf.\ Ser.~267, Hot Star Workshop III: The Earliest Stages of Massive Star Birth, ed.\ P.A.\ Crowther (San Francisco:  ASP), 111

\bibitem[]{} Walborn, N.R., Barb\'a, R.H., Brandner, W., Rubio, M., Grebel, E.K., \& Probst, R.G. 1999, \aj, 117, 225

\bibitem[]{} Walborn, N.R., \& Blades, J.C. 1997, \apjs, 112, 457

\bibitem[]{} Walborn, N.R., Ma\'{\i}z-Apell\'aniz, J., \& Barb\'a, R.H.
2002, \aj, 124, 1601

\bibitem[]{} Walborn, N.R., \& Parker, J.Wm. 1992, \apj, 399, L87

\bibitem[]{} Werner, M.W., Becklin, E.E., Gatley, I., Ellis, M.J., Hyland, A.R., Robinson, G., \& Thomas, J.A. 1978, \mnras, 184, 365

\bibitem[]{} Whitney, B.A., Sewi{\l}o, M., Indebetouw, R., Robitaille, T.P., 
Meixner, M., Gordon, K. et al. 2008, \aj, 136, 18

\bibitem[]{} Woods, P.M. et al. 2011, \mnras, 411, 1597

\end{thebibliography}
\end{document}